# Agent-Based Model for Rural-Urban Migration: A Dynamic Consideration

Cai, Ning[1]   Ma, Hai-Ying[2]   Khan, M. Junaid[3]

[1]College of Electrical Engineering, Northwest University for Nationalities, Lanzhou, China
[2]College of Economics, Northwest University for Nationalities, Lanzhou, China
[3]PN Engineering College, National University of Sciences and Technology, Karachi, Pakistan

**Abstract:** This paper develops a dynamic agent-based model for rural-urban migration, based on the previous relevant works. The model conforms to the typical dynamic linear multi-agent systems model concerned extensively in systems science, in which the communication network is formulated as a digraph. Simulations reveal that consensus of certain variable could be harmful to the overall stability and should be avoided.

**Key Words:** Rural-Urban Migration; Multi-Agent Systems; Digraph; Consensus

## 1. Introduction

The economic growth in developing countries almost inevitably leads to substantial flow of the labor force between rural agriculture and industries in cities [1]. Many economic models have appeared since 1950s to expound the population migrations. Among them, the most famous and successful model was developed by Harris and Todaro [2], paying attention to the expected income differentials between the two sectors.

There exist inconsistencies from the interpretation by traditional economic theory to the reality of population migration. Possibly, this is in part because that the real labor market is a sophisticated large-scale dynamic system, with vast feedbacks and emergencies, and it can hardly be depicted by any simple structure. The macroscopic dynamics of economics are formed from the local interactions of individuals, whereas the macroscopic dynamics would in turn impact the microscopic behaviors of individuals.

Since the last decade, agent-based computational economics (ACE) have aroused extensive interest from the academic community, which is based on a new bottom-up modeling methodology. Such a methodology actually integrates the notion of complex dynamic systems into economics. So far, several articles have arisen in the literature being relational to the study of rural-urban migration with ACE methods. For instance, Van Dyke Parunak *et al.* [3] earlier discussed agent-based modeling of rural-urban migration. Tesfatsion [4] conducted experiments to study the hysteresis effect from communication networks. Silveira *et al.* [5] built a statistical mechanics model based on probability theory and ACE methods. Itoh [6] developed a scheme for the dynamic control of urbanization. Kenivetona *et al.* [7] endeavored to predict population migration with considering the environmental influences. Filho *et al.* [8] proposed an evolutionary agent-based schema to simulate migrations, concerning the possibly high-order dynamics. Most of these results are empirical, lacking in-depth theoretical analysis.

---

[1] Corresponding author: Cai, Ning (e-mail: caining91@tsinghua.org.cn)

The current paper is primarily devoted to extend the layout in [5] via developing a dynamic multi-agent model for the sake of simulating the rural-urban migrations.

The motivation of this work partially emerges from our explorations [9-13] of theory on dynamic multi-agent systems in systems science, especially the consensus problem. Consensus of a system implicates asymptotic stability in certain information, and it actually plays the role as equilibrium of isolated dynamical systems [9-10]. Due to its theoretical significance, the consensus problem has received extensive concern from various perspectives; see [9-21] and references therein. Particularly, a general framework of the consensus problem for first-order systems is initially proposed in [15]. A necessary and sufficient condition for consensus of LTI first-order systems is given in [16]. Consensus problem of high-order systems is earlier addressed in [17-18]. Both observer-based dynamical protocol and robust consensus are handled in [19]. A noteworthy approach about oblique decomposition of the state space is devised in [13, 20]. A method based on relative Lyapunov function is proposed in [12], which could potentially handle the consensus problems of systems with nonlinear or even heterogeneous agents.

The theories about dynamic multi-agent systems have been highly developed in systems science. However, the application instances associated with these theories that can well support them are seriously absent. Our attempt introduces a scenario from economics, in order to apply, verify, and enrich the relevant theories in systems science.

As far as our knowledge is concerned, nearly all the existing discussions seek an achievement of consensus, without considering any non-consensus situation; whereas our research endeavors oppositely to turn the viewpoints to the anti-consensus cases by revealing that sometimes consensus maybe detrimental.

The advantages of the current work are evident in four aspects, comparing with [5]. 1) Any real economic system should be dynamic. But the model in [5] is static without dynamics. 2) The article [5] brought unnecessary nonlinearities into the migratory mechanism, whereas our model keeps being linear, conforming to the typical state space framework to describe dynamic systems. This would be superior in pursuing the principles of migration analytically, with the potential utilization of the abundant theories from systems science. For instance, a precise condition for consensus can be derived according to the present model. 3) The migratory process in [5] is iterative, whereas the variables in the current model, such as the time and the intention to migrate, are continuous. This is evidently closer to the reality. 4) The current model takes the influence from the topology of communication network into account.

The organization of the main body of this paper is as follows. Section 2 briefly introduces the basic economic setting as background knowledge. Section 3 describes the dynamic migration mechanism in detail. Section 4 discusses the relevant consensus problem, analytically. Section 5 contains several simulation instances to illustrate the model. Finally, Section 6 concludes the paper.

## 2. Economic Setting

The economy of a society is supposed to comprise two sectors: urban and rural. The urban sector is formed by firms engaged in manufacturing business, whilst the rural sector is formed by farms engaged in agriculture. Such a dual structure is typically adopted by the economic literature addressing the rural-urban migration phenomena [1-2, 5, 22-23].

The basic economic setting of both sectors in the current paper is inherited from [5]. The primary conclusions are quoted as follows. The readers can refer to [5] for detailed analysis.

Let $N_u$ denote the population of urban sector. The average wage in the manufacturing sector is a function of $N_u$

$$w_m = \xi_2 N_u^{\alpha-1} \qquad (1)$$

where $0 < \alpha < 1$ and $\xi_2$ is a constant calculated as

$$\xi_2 = \alpha A_m \left(\frac{\eta}{1-\eta}\right)^{\alpha\eta} \left[\left(1-\frac{\eta}{b}\right)\frac{1}{Z_m}\right]^{\alpha-1}$$

with the parametric constants $A_m > 0$, $0 < \eta < 1$, $b > 0$, and $Z_m$ denoting the quantity of firms.

In the rural sector, the average income per worker is also a function of the rural population

$$w_a = \xi_4 p(N - N_u)^{\phi-1} \tag{2}$$

where $0 < \phi < 1$; $N$ denotes the population of entire society; and $\xi_4$ is a constant calculated as

$$\xi_4 = A_a \phi / Z_a^{\phi-1}$$

with the parametric constants $A_a > 0$ and $Z_a$ denoting the quantity of farms.

In (2), $p$ is the relative price per unit of agricultural goods, which is calculated as

$$p = \rho (Y_m / Y_a)^\gamma$$

where $\rho, \gamma > 0$ are parametric constants and $Y_m$ and $Y_a$ denote the aggregated productions of the manufacture sector and agriculture sector, respectively. $Y_m$ is calculated as

$$Y_m = \xi_1 N_u^\alpha$$

with

$$\xi_1 = A_m Z_m^{1-\alpha} \left[\left(\frac{\eta}{1-\eta}\right)^\eta \left(1-\frac{\eta}{b}\right)\right]^\alpha$$

Meanwhile, $Y_a$ is calculated as

$$Y_a = \xi_3 (N - N_u)^\phi$$

with

$$\xi_3 = A_a Z_a^{1-\phi}$$

## 3. Migratory Process: A Dynamic Agent-Based Consideration

The dynamic system of equations is as follows:

$$\begin{cases} \dot{x}_1 = ax_1 + f\sum_1^N w_{j1}(x_j - x_1) + bv \\ \dot{x}_2 = ax_2 + f\sum_1^N w_{j2}(x_j - x_2) + bv \\ \vdots \\ \dot{x}_N = ax_N + f\sum_1^N w_{Nj}(x_j - x_N) + bv \end{cases} \tag{3}$$

In this system of equations, the state variable $x_i \in R$ measures the intention of worker $i$ to work in the rural or urban sector. $x_i > 0$ implies that the worker intends to work in urban sector, whereas $x_i < 0$ implies an intention to work in rural sector. The greater the absolute value of $x_i$, the stronger this intention is. The unit of time $t$ is a day.

The intention dynamics of each worker is jointly determined by three components.

The first component $ax_i$ reflects the inertia of worker $i$. If a worker stays in one sector for a long time, he inclines to get used to the current status quo and keep on staying. Thus, the coefficient $a \in R^+$.

The second component $f\sum_{1}^{N} w_{ij}(x_j - x_i)$ represents the social influence from one's companions. The communication architecture of the society is represented by a graph $G$ of order $N$, with each worker corresponding to a vertex. The arc weight of $G$ from vertex $j$ to $i$ is denoted by $w_{ij} \geq 0$, which can be regarded as the strength of information link. If $w_{ij} > 0$, this means that vertex $j$ is a neighbor of vertex $i$. The graph $G$ is directed, which can be denoted by its adjacency matrix $W$:

$$G:W = \begin{bmatrix} w_{11} & w_{12} & \cdots & w_{1N} \\ w_{21} & w_{22} & \cdots & w_{2N} \\ \vdots & \vdots & & \vdots \\ w_{N1} & w_{N2} & \cdots & w_{NN} \end{bmatrix}$$

People usually tend to think their friends are wiser than themselves and then follow others, especially in Asian countries such as China. Thus, the social influence should be positive, with the coefficient $f \in R^+$.

The last component $bv$ stands for one's rational judgment, according to the actual economic situation. The expected wage differential between urban and rural sectors can be regarded as the input variable $v$ of system (3), i.e.

$$v = (1 - r_u)w_m - w_a$$

where $r_u$ denotes the unemployment rate in the urban sector. Evidently, the input coefficient $b \in R^+$.

Every month, each worker reviews the situation and decides whether to migrate or stay, respectively. A worker is potentially to migrant if his intention to work in the other sector is high. For worker $i$, an effective migration will take place with probability

$$P_i = \frac{|x_i|}{|x_i| + \beta} \tag{4}$$

The personal idiosyncratic motivation, or random private utility [5, 22], is modeled by the above equation. It is worth mentioning that the fraction in (4) can well map from the domain $[0, +\infty)$ or $(-\infty, 0]$ to a domain $[0,1)$. $\beta \in R^+$ is a parameter to modulate the sensitivity. Migration is more sensitive with a smaller $\beta$.

The state space equation (3) takes the form of the linear multi-agent system that has been studied profoundly in systems science. A major benefit of this kind of formulation is that not only experimentally simulative, but also analytic results about rural-urban migration could possibly be derived with the aid of the existing theories on multi-agent systems. In the next section, this will be exemplified by analyzing the consensus problem.

## 4. Consensus Analysis

*Definition 1:* (**Consensus**) For the dynamic system (3), if
$$\lim_{t \to \infty}(x_i - x_j) = 0 \quad (\forall i, j \in \{1, 2, ..., N\})$$
then the system achieves *consensus*.

*Definition 2* [24]: (**Laplacian Matrix**) The *Laplacian matrix* of a weighted directed graph can be computed as
$$L = D - W$$
where $D$ is the in-degree matrix and $W$ is the adjacency matrix.

*Lemma 1* [16]: The Laplacian matrix $L$ of a directed graph $G$ has exactly a single zero eigenvalue $\lambda_1 = 0$ if and only if $G$ has a spanning tree, with the corresponding eigenvector $\phi = \begin{bmatrix} 1 & 1 & \cdots & 1 \end{bmatrix}^T$. Meanwhile, all the other eigenvalues $\lambda_2, ..., \lambda_N$ locate in the open right half plane.

*Lemma 2* [9-11]: The autonomous dynamic system
$$\dot{x}_i = ax_i + f \sum_{1}^{N} w_{1j}(x_j - x_i) \quad (i = 1, 2, ..., N)$$
with $a > 0$ achieves consensus if and only if both 1) and 2) below are true:
1) The graph topology $G$ includes a spanning tree;
2) All the values $a - \lambda_i f$ ($i \in \{1, 2, ..., N\}$ $\lambda_i \neq 0$) have negative real parts.

where $\lambda_1, \lambda_2, ..., \lambda_N$ are the eigenvalues of the Laplacian matrix.

*Proposition 1:* The condition for the forced system (3) to achieve consensus is the same with Lemma 2.

*Proof:* According to Definition 1, consensus is determined by the dynamics of all the relative states $x_i - x_j$ ($\forall i, j \in \{1, 2, ..., N\}$), which can be derived by subtracting between different equations in (3). These dynamics are independent of the identical input term *bv*. □

Consensus as a topic has been extensively studied in systems science. It essentially equals a kind of asymptotic stability [10-11]. Usually, instability is regarded to be meaningless in the community of systems science. Almost all papers in the literature concerning the topic of consensus only pay attention to the case that consensus would occur. However, in the scenario of the current paper, a society-wide consensus of intentions is apparently detrimental. What we really expect is an anti-consensus mode.

*Remark 1:* Consensus of certain variable could in turn result in the overall instability of a system. Actually such a phenomenon is usual in economic areas, besides rural-urban migration, e.g., in the fields of investment such as the stock market.

The graph topology of any society surely has a spanning tree. As a corollary of Proposition 1 and Lemma 1, the condition of anti-consensus can be summarized as follows.

*Proposition 2:* The intention dynamics of (3) can avoid consensus if and only if
$$a > f \operatorname{Re}(\lambda_2)$$
where $\lambda_2$ is the nonzero eigenvalue of Laplacian matrix that has the minimal real part.

*Remark 2:* In order to avoid consensus, *a* should be sufficiently large, otherwise *f* or $\operatorname{Re}(\lambda_2)$ should be small enough. If $a \leq 0$, then consensus is inevitable. In other words, the private inclination should overwhelm the social influence.

*Remark 3:* It might be worth mentioning that the magnitude of the spectrum of a graph topology is proportional to its average weight of arcs. Therefore, consensus is prone to happen if with a higher average strength of links in the communication network.

## 5. Simulations

The framework of simulation is similar to that in [5]. For convenience of scaling, each worker is placed at one site of a square lattice, in order to visually illustrate the status of them. Notice that this is only a way of arrangement and the coordinates of the lattice sites are not related to the actual geographical locations of workers.

At the initial stage, the social graph topology is generated randomly. First, let $w_{ij}$ be random numbers uniformly distributed in the interval (0, 0.1). Then, in order to reduce the average degree of graph, set those elements which are less than a given threshold be zero. Such a threshold is called sparse factor here.

A social graph topology of order 100 is illustrated in Fig. 1. Notice that in this figure, the arcs are actually directed, with the arrows omitted.

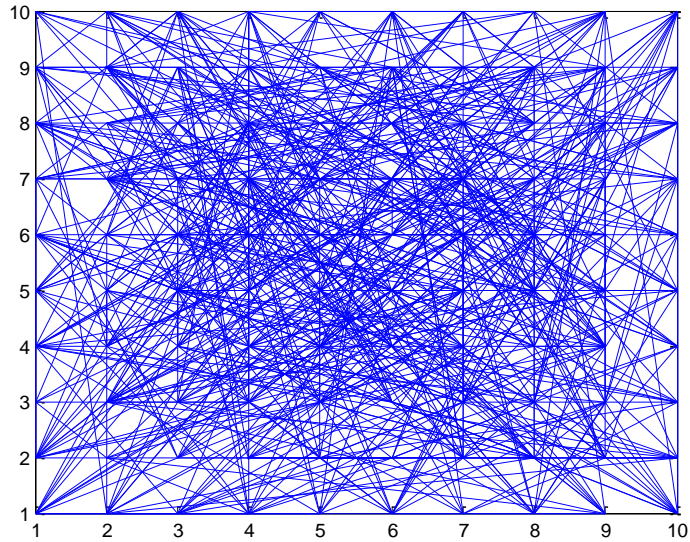

Fig. 1. A graph with sparse factor 0.095

In the first instance, let $a = 0.0008$, $f = 0.001$, $b = 0.02$, and $\beta = 3$. The sparse factor of the graph topology is set as 0.09. The initial ratio of urban population is set as 20%, which is typical among pre-urbanized developing countries. According to Proposition 2, no consensus will occur in this case.

The initial distribution and the distribution after 30 months are shown in Figures 2 & 3, respectively. In addition, the transition curves of the urban population $N_u$ and the value of input term $bv$ are shown in Figures 4 & 5. One can see that although the balance is stable, the motion has overshoot.

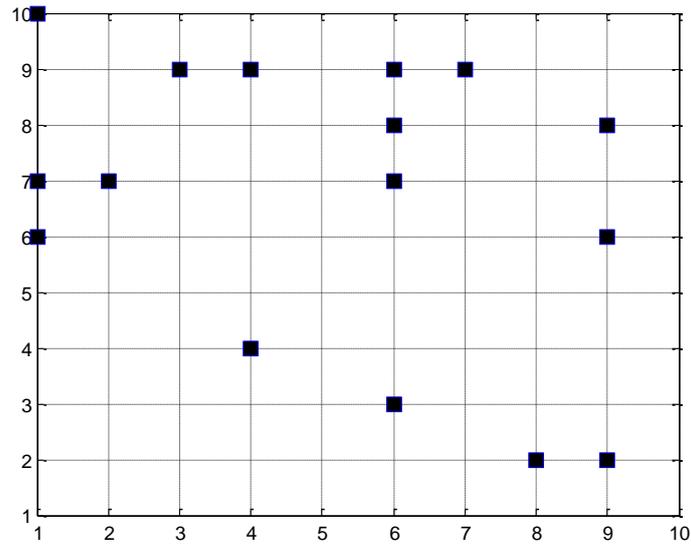

Fig. 2. Initial distribution
Thick dots denote workers in urban sector.

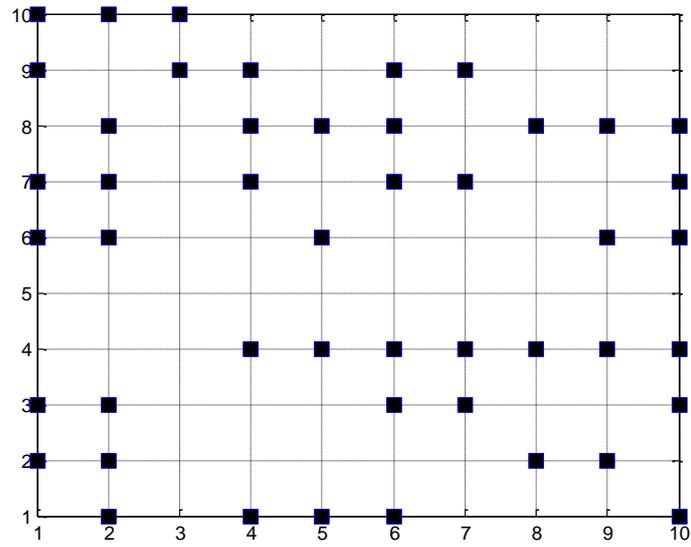

Fig. 3. Distribution after 30 months

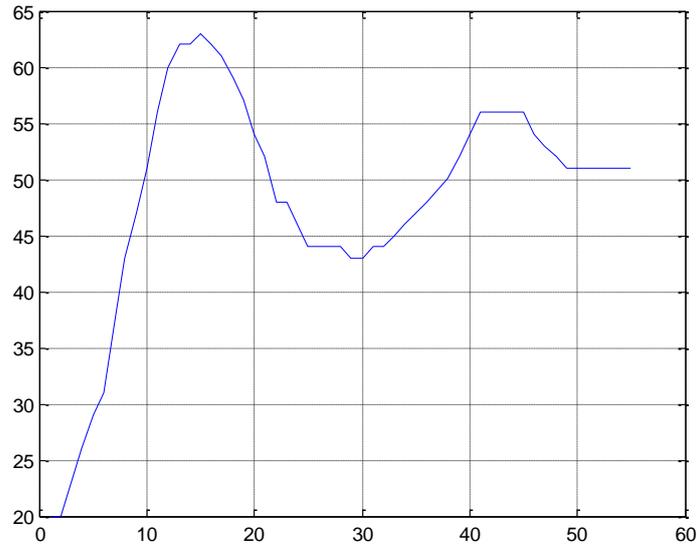

Fig. 4. Transition of urban sector population $N_u$ in anti-consensus case

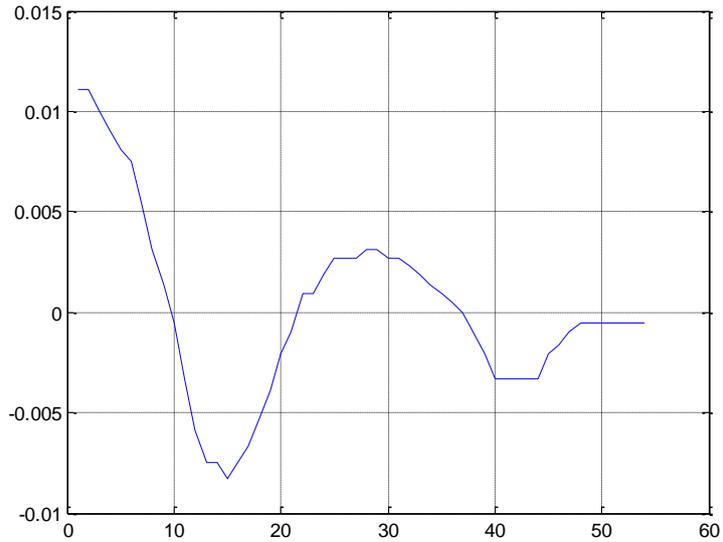

Fig. 5. Transition of input term $bv$ in anti-consensus case

In the second instance, let $a = 0.002$, $f = 0.004$, $b = 0.02$, and $\beta = 2$. The sparse factor of the graph topology is set as 0.08. This time consensus will occur. From the transition curve of the urban population $N_u$ shown in Fig. 6, one can see that there is no convergence to the balance.

In China, there is a household registration system called "hukou" system, regulated by the government. Under this system, a portion of urban residents have citizen hukou, which is a kind of registered permanent residence. Most of them have lived in cities for generations. Usually these people would not migrate to the countryside for a living even if their actual intention is intensive.

Assume that the 20% people who initially live in cities hold citizen hukou. As a simulating result, the transition curve of the urban population $N_u$ is shown in Fig. 7. In comparison with Fig. 6, the oscillation is notably suppressed. In this sense, the household registration system is beneficial.

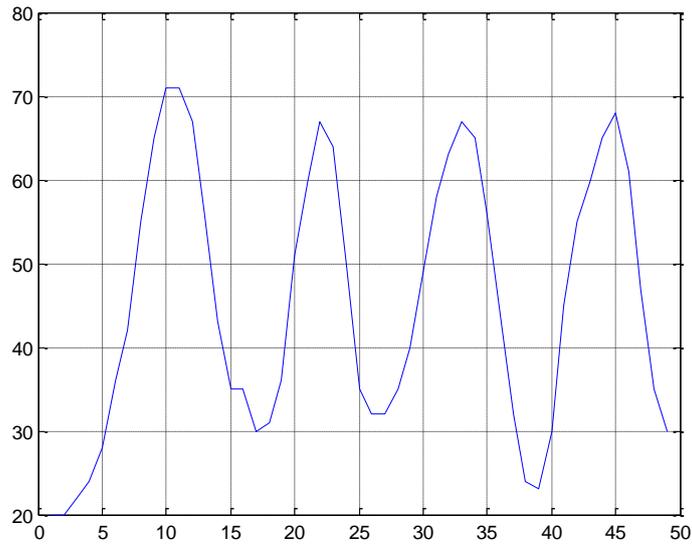

Fig. 6. Transition of urban sector population $N_u$ in consensus case without household registration system

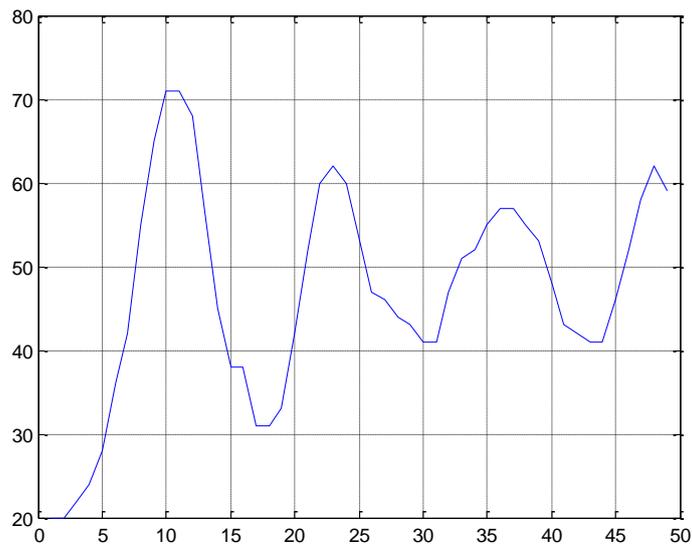

Fig. 7. Transition of urban sector population $N_u$ in consensus case with household registration system

# 6. Conclusion

This paper extends the previous relevant works in the literature and proposes a model for rural-urban migration, which is not only agent-based, but also dynamic. According to this model, the motion of the individual intentions to migrate can be formulated by a system of linear differential equations, with the equations interacting across a communication network represented by a directed graph topology. Such a dynamic model is compatible with the typical form of multi-agent systems studied in systems science. Thus, the ample existing theories are convenient to be exploited in analyzing the principles of rural-urban migration, and reversely, the economic phenomena could help to stimulate the new explorations in systems science. As an instance, our research indicates that consensus should be avoided because it may damage the overall stability. In the future, a lot of new academic tasks can be conducted along the current technical route, both in economics and systems science. For example, simulations can be progressively carried out under various parametric settings; also, the theoretical mechanism of the anti-consensus mode of general dynamic multi-agent systems can be probed in depth.

# Acknowledgements

This work is supported by Program for Young Talents of State Ethnic Affairs Commission (SEAC) China (Grant [2013] 231), and by National Natural Science Foundation (NNSF) of China (Grants 61174067, 61263002, & 61374054).

# References

[1] D. Ray, *Development Economics*, Princeton University Press: Princeton, 1998.

[2] J.R. Harris & M.P. Todaro, "Migration, unemployment and development: A two-sector analysis", *Am. Econ. Rev.*, vol. 60, no. 1, pp. 126-142, 1970.

[3] H. Van Dyke Parunak, R. Savit, & R.L. Riolo, "Agent-based modeling vs. equation-based modeling: A case study and users' guide", *Lecture Notes in Computer Science*, vol. 1534, pp. 277-283, 1998.

[4] L. Tesfatsion, "Introduction to the special issue on agent-based computational economics", *J. Econ. Dyn. Control*, vol. 25, no. 3-4, pp. 281-293, 2001.

[5] J.J. Silveira, A.L. Espindola, & T.J.P. Penna, "Agent-based model to rural-urban migration

analysis", *Physica A*, vol.264, pp.445-456, 2006.

[6] R. Itoh, "Dynamic control of rural-urban migration", *J. Urban Econ.*, vol. 66, no. 3, pp. 196-202, 2009.

[7] D. Knivetona, C. Smitha, & S. Wood, "Agent-based model simulations of future changes in migration flows for Burkina Faso", *Global Environmental Change*, vol. 21, pp. S34-S40, 2011.

[8] H.S.B. Filho, F.B.L. Neto, & W. Fusco, "Migration and social networks - An explanatory multi-evolutionary agent-based model", *Proc. IEEE Symposium on Intelligent Agent*, pp. 1-7, 2011.

[9] N. Cai, *Swarm Stability and Controllability of High-Order Swarm Systems*, Doctoral Dissertation, Tsinghua University, 2010. (In Chinese)

[10] N. Cai, J. Xi, & Y. Zhong, "Swarm stability of high order linear time-invariant swarm systems", *IET Control Theory Appl.*, vol. 5, no. 2, pp. 402-408, 2011.

[11] N. Cai, J. Xi, & Y. Zhong, "Asymptotic swarm stability of high order dynamical multi-agent systems: Condition and application", *Control and Intelligent Systems*, vol. 40, no. 1, pp. 33-39, 2012.

[12] N. Cai, J. Cao, & H. Ma *et al.*, "Swarm stability analysis of nonlinear dynamical multi-agent systems via relative Lyapunov function", *Arab. J. Sci. Eng.*, vol. 39, no. 3, pp. 2427-2434, 2014.

[13] J. Xi, N. Cai, & Y. Zhong, "Consensus problems for high-order linear time-invariant swarm systems", *Physica A*, 2010, vol. 389, no. 24, pp. 5619-5627.

[14] A. Jadbabaie, J. Lin, & A. Morse, "Coordination of groups of mobile autonomous agents using nearest neighbor rules", *IEEE Trans. Autom. Control*, vol. 48, no. 6, pp. 988-1001, 2003.

[15] R. Olfati-Saber & R. Murray, "Consensus problems in networks of agents with switching topology and time-delays", *IEEE Trans. Autom. Control*, vol. 49, no. 9, pp. 1520-1533, 2004.

[16] W. Ren & R. Beard, "Consensus seeking in multiagent systems under dynamically changing interaction topologies", *IEEE Trans. Autom. Control*, vol. 50, no. 5, pp. 655-661, 2005.

[17] F. Xiao & L. Wang, "Consensus problems for high-dimensional multi-agent systems", *IET Control Theory Appl.*, vol. 1, no. 3, pp. 830-837, 2007.

[18] J. Wang, D. Cheng, & X. Hu, "Consensus of multi-agent linear dynamic systems", *Asian J. Control*, vol. 10, no. 2, pp. 144-155, 2008.

[19] Z. Li, Z. Duan, & G. Chen *et al.*, "Consensus of multiagent systems and synchronization of complex networks: A unified viewpoint", *IEEE Trans Circuit. Syst. I: Regular Papers*, vol. 57, no. 1, pp. 213-224, 2010.

[20] J. Xi, Z. Shi, & Y. Zhong, "Consensus analysis and design for high-order linear swarm systems with time-varying delays", *Physica A*, vol. 390, no. 23-24, pp. 4114-4123, 2011.

[21] H. Hu, W. Yu, & Q. Xuan *et al.*, "Consensus for second-order agent dynamics with velocity estimators via pinning control", *IET Control Theory Appl.*, vol. 7, no. 9, pp. 1196-1205, 2013.

[22] L.H. Summers, "Relative wages, efficiency wages, and Keynesian unemployment", *Am. Econ. Rev.*, vol. 78, no. 2, pp. 383-388, 1988.

[23] W.A. Brock & S.N. Durlauf, "Discrete choice with social interactions", *Rev. Econ. Stud.*, vol. 68, no. 2, pp. 235-260, 2001.

[24] C. Godsil & G. Royle, *Algebraic Graph Theory*, Springer-Verlag: Berlin, 2001.